\documentclass[10pt,conference]{IEEEtran}
\usepackage[inkscapelatex=false]{svg}
\IEEEoverridecommandlockouts
\usepackage{cite}
\usepackage{amsmath,amssymb,amsfonts}
\usepackage{graphicx}
\usepackage{textcomp}
\usepackage{amsmath}
\usepackage{booktabs} 
\usepackage{multirow}
\usepackage{hyperref}       
\usepackage{url}            
\usepackage{booktabs}       
\usepackage{amsfonts}       
\usepackage{nicefrac}       
\usepackage{microtype}      
\usepackage{xcolor}         
\usepackage{algorithm}
\usepackage{algpseudocode}
\usepackage{qcircuit}
\usepackage{amsmath}
\usepackage{tikz}
\usepackage{bbm}

\usepackage{orcidlink}
\usepackage{comment}
\usepackage{amssymb}
\usepackage{xcolor}
\def\BibTeX{{\rm B\kern-.05em{\sc i\kern-.025em b}\kern-.08em
    T\kern-.1667em\lower.7ex\hbox{E}\kern-.125emX}}
\definecolor{tensorpurp}{rgb}{1,0.5,1}
\tikzset{w/.style={fill=green!50!black!80,draw=black}}
\tikzset{ten/.style={fill=tensorpurp}}

\newcommand{\ket}[1]{|{#1}\rangle}

\newcommand{\diagram}[1]{
    \begin{array}{cc}
        \begin{tikzpicture}[scale=.5,every node/.style={sloped,allow upside down},
        baseline={([yshift=+0ex]current bounding box.center)}]
            #1
        \end{tikzpicture}
    \end{array}
}

\newcommand{\del}[1]{}

\begin{document}

\title{QAS-QTNs: Curriculum Reinforcement Learning-Driven Quantum Architecture Search for Quantum Tensor Networks}

\author{\IEEEauthorblockN{Siddhant Dutta\textsuperscript{1}, Nouhaila Innan\textsuperscript{2,3}, Sadok Ben Yahia\textsuperscript{4,5}, Muhammad Shafique\textsuperscript{2,3}
}

\IEEEauthorblockA{\textsuperscript{1}College of Computing and Data Science, Nanyang Technological University, Singapore\\
\textsuperscript{2}eBRAIN Lab, Division of Engineering, New York University Abu Dhabi (NYUAD), Abu Dhabi, UAE\\
\textsuperscript{3}Center for Quantum and Topological Systems (CQTS), NYUAD Research Institute, NYUAD, Abu Dhabi, UAE\\
\textsuperscript{4}The Maersk Mc-Kinney Moller Institute, University of Southern Denmark, Sønderborg, Denmark\\
\textsuperscript{5}Corvinus Institute for Advanced Studies (CIAS), Budapest, Hungary\\ 
\vspace{0.2cm}
\small siddhant010@e.ntu.edu.sg, nouhaila.innan@nyu.edu, say@mmmi.sdu.dk,
muhammad.shafique@nyu.edu
}}

\maketitle

\begin{abstract}
Quantum Architecture Search (QAS) is an emerging field aimed at automating the design of quantum circuits for optimal performance. This paper introduces a novel QAS framework employing hybrid quantum reinforcement learning with quantum curriculum learning strategies, enabling learning agents to tackle increasingly complex quantum circuit design tasks. We benchmark four state-of-the-art classical reinforcement learning algorithms (A2C, PPO, DDQN, TD3) against their quantum-enhanced counterparts (QA2C, QPPO, QDDQN, QTD3) for optimizing variational quantum circuits (VQCs). Our approach progressively increases circuit depth and gate complexity during training, leveraging parameterized quantum circuits as function approximations. To improve learning efficiency and stability, all algorithms, both classical and quantum, are augmented with Prioritized Experience Replay (PER). Experimental results show that quantum-enhanced RL significantly outperforms classical methods. In a 2-qubit environment, PERQDDQN achieves a success probability of 0.46 with ~3,000 optimal successes, surpassing classical PERDDQN (0.42, ~2,400). In the more complex 3-qubit setting, PERQDDQN and PERQTD3 reach success probabilities of ~0.47, with optimal success counts of ~3,800 and ~3,600, respectively, outperforming their classical counterparts. Additionally, we apply our QAS-QTN approach to a classification problem, where the optimized quantum circuit achieves an accuracy of 90.33\%, outperforming quantum models consisting of random ansatz. This hybrid classical-quantum approach leads to faster convergence and more efficient quantum circuit designs, demonstrating its potential for advancing automated quantum architecture search.
\end{abstract}


\section{Introduction}

Advances in variational quantum computing underscore the need for effective methods to design and optimize quantum circuits, as circuit performance directly impacts the success of quantum algorithms. Quantum Architecture Search (QAS) addresses this challenge by automating the search for optimal quantum circuit designs \cite{du2022quantum,kuo2021quantum}. Traditional methods, such as manual or heuristic approaches, often fail to scale with the increasing complexity of quantum systems. Leveraging automated techniques like Deep Reinforcement Learning (DRL), QAS explores vast search spaces to enhance circuit designs for emerging quantum hardware.

Quantum Tensor Networks (QTNs) have emerged as a powerful tool in quantum computing \cite{haghshenas2022variational}, particularly for applications in quantum machine learning. QTNs offer an efficient way to represent and manipulate high-dimensional quantum states by decomposing complex states into simpler, lower-dimensional tensors. This decomposition makes it easier to simulate and compute quantum systems, as tensor networks can capture the entanglement structure and correlations inherent in quantum states—essential elements for efficient computation. With tensor networks, any quantum state can be represented as a network of tensors, where the bond dimension controls the expressibility and complexity of the state \cite{biamonte2019lectures}. One of the key features of QTNs is their ability to represent any quantum state as a tensor network, with the bond dimension controlling the complexity and expressibility of the state. These tensor networks like Matrix Product State (MPS) and Tree-like Tensor Network (TTN) break down a quantum state into smaller, more manageable components, making it easier to simulate and manipulate, particularly for high-dimensional or strongly correlated systems \cite{ORUS2014117,Huggins_2019}.

DRL has become a cornerstone technique in optimizing quantum circuits, particularly in the context of Quantum Architecture Search \cite{fosel2021quantum}. In DRL, an agent learns to make decisions through trial and error, receiving feedback in the form of rewards or penalties based on its actions. In the quantum domain, this is used to train models that improve quantum circuit designs iteratively. Classical RL techniques such as Q-learning, Actor-Critic, and Proximal Policy Optimization (PPO) have been adapted for quantum problems, where the state space is typically vast, and the environment is governed by quantum mechanical principles \cite{zhu2023quantum, liu2023quantum}. DRL-based automation helps design quantum circuits by guiding the search for optimal circuit structures that minimize error rates and maximize fidelity in quantum computations \cite{ostaszewski2021reinforcement}. The advantage of DRL lies in its ability to adapt to complex quantum systems by continually improving circuit architectures based on feedback, making it particularly suitable for tackling the challenges posed by Noisy Intermediate-Scale Quantum (NISQ) devices \cite{preskill2018quantum}.

Hybrid Quantum Reinforcement Learning (QRL) with quantum techniques has recently shown the potential to improve the efficiency and performance of the agent's optimal decision-making capability. These approaches often involve adapting classical Reinforcement Learning (RL) methods like Double Deep Q-Network (DDQN) to quantum systems \cite{dutta2024qadqn}. This capability is further enhanced when coupled with quantum curriculum learning, where the complexity of the quantum tasks gradually increases, allowing the learning agent to build progressively more sophisticated circuit designs \cite{tran2024quantum}.

In this paper, we address the challenges of optimizing quantum circuits using QAS and propose a novel framework that integrates DRL and quantum curriculum learning. We focus on automating the design of Variational Quantum Circuits (VQCs) using QTNs, where traditional circuit design methods struggle with scalability and complexity, particularly in NISQ devices. Our contributions are:

\begin{itemize} 
    \item We propose the QAS-QTN framework that automates the design of quantum circuits by leveraging quantum deep reinforcement learning and quantum curriculum learning. This framework significantly reduces manual effort in the design process by systematically exploring design spaces. 
    \item We evaluate several hybrid classical-quantum approaches that integrate classical RL algorithms with quantum-enhanced variants to optimize quantum circuit architectures, aiming to enhance their performance on NISQ devices.
    \item We automate the incremental learning process through quantum curriculum learning, gradually increasing the task complexity, and allowing the model to handle increasingly challenging quantum circuit optimization tasks.  
    \item We demonstrate the effectiveness of our quantum approach by applying it to a real-world dataset, where the QAS-QTN optimized quantum circuit model achieves an accuracy of 90.33\%, outperforming quantum models consisting of random ansatz.
\end{itemize}

The structure of the paper is as follows: Sec. \ref{se2} provides the preliminaries and reviews related work. Sec. \ref{se3} outlines the problem formulation and methodology in Sec. \ref{se31}, detailing the Curriculum Learning-based QAS-QTN framework in Sec. \ref{se32}. In Sec. \ref{se4}, the experimental setup is presented in  Sec. \ref{se41}, and the simulation results with comparisons are discussed in Sec. \ref{se42}. Finally, Sec. \ref{se5} offers concluding remarks and discusses potential open challenges for future research.

\section{Preliminaries and Related Work \label{se2}}

\subsection{Quantum Architecture Search}

QAS aims to automate the design of quantum circuits through optimization techniques. Early work primarily employed classical optimization methods, such as Genetic Algorithms (GA) and Simulated Annealing (SA), to search through the quantum circuit design space \cite{lu2023qas}. These methods optimize the parameterized quantum circuit by minimizing a loss function defined as:
\begin{equation}
    \mathcal{L}(\theta) = \| U(\theta) |\psi\rangle - |\psi_{\text{target}}\rangle \|^2,
\end{equation}
where \( U(\theta) \) represents the parameterized quantum circuit, \( |\psi\rangle \) is the quantum state produced by the circuit, and \( |\psi_{\text{target}}\rangle \) is the target state. This loss function quantifies the difference between the quantum state and the target state, which is minimized during optimization.

\subsection{Reinforcement Learning}

In recent years, RL has been applied to QAS, where the quantum circuit design is viewed as a sequential decision-making problem. In this approach, the agent learns to optimize quantum circuit parameters by maximizing a reward function based on the performance of the quantum circuit \cite{sun2023differentiable,altmann2023challenges}. The reward function can be expressed as:
\begin{equation}
R(\theta) = \text{Tr} \left( \rho(\theta) \hat{O} \right),
\end{equation}
where \( \rho(\theta) \) is the quantum state produced by the quantum circuit \( U(\theta) \) with parameters \( \theta \), and \( \hat{O} \) is an observable that evaluates the quantum circuit's performance, such as fidelity or expectation values.

In the context of DRL, the agent learns a policy \( \pi(a|s) \) to map states \( s \) to actions \( a \), optimizing the expected future reward. The Q-value function, which estimates the expected reward for each state-action pair, is defined as:
\begin{equation}
Q(s, a) = \mathbb{E}\left[ \sum_{t=0}^{T} \gamma^t R(s_t, a_t) \mid s_0 = s, a_0 = a \right],
\end{equation}
where \( \gamma \) is the discount factor that prioritizes immediate rewards \cite{mnih2013playing}. The Q-value function is updated iteratively using Bellman's Equation:
\begin{equation}
Q(s, a) \leftarrow Q(s, a) + \alpha \left( R(s, a) + \gamma \max_{a'} Q(s', a') - Q(s, a) \right),
\end{equation}
where \( \alpha \) is the learning rate, and \( s' \) is the next state after taking action \( a \) in state \( s \) \cite{o2018uncertainty}.

PPO is a policy gradient method designed to improve the stability and efficiency of training by clipping the probability ratio between the old and new policies. The objective function for PPO is:
\begin{equation}
L_{\text{PPO}}(\theta) = \mathbb{E}_t \left[ \min \left( r_t(\theta) \hat{A}_t, \text{clip}(r_t(\theta), 1 - \epsilon, 1 + \epsilon) \hat{A}_t \right) \right],
\end{equation}
where \( r_t(\theta) = \frac{\pi_{\theta}(a_t|s_t)}{\pi_{\theta_{\text{old}}}(a_t|s_t)} \) is the probability ratio of the new and old policy, \( \hat{A}_t \) is the estimated advantage function at time step \( t \), and \( \epsilon \) is a small clipping parameter to prevent large updates to the policy \cite{schulman2017proximal}.

The advantage function \( \hat{A}_t \) is typically estimated using Generalized Advantage Estimation (GAE):
\begin{equation}
    \hat{A}_t = \delta_t + (\gamma \lambda) \delta_{t+1} + (\gamma \lambda)^2 \delta_{t+2} + \cdots,
\end{equation}
where \( \delta_t \) is the temporal difference error:
\begin{equation}
\delta_t = R_t + \gamma V(s_{t+1}) - V(s_t).
\end{equation}

A2C is an actor-critic method where the policy (the actor) and the value function (the critic) are learned simultaneously. It is also shown to be a special case of PPO \cite{huang2022a2c}. The objective function is composed of two terms: the policy gradient and the value function loss. The actor's objective is to maximize the expected advantage:
\begin{equation}
L_{\text{actor}}(\theta) = \mathbb{E}_t \left[ \nabla_{\theta} \log \pi_{\theta}(a_t|s_t) \hat{A}_t \right].
\end{equation}

The critic minimizes the mean squared error between the predicted value and the target value:
\begin{equation}
L_{\text{critic}}(\phi) = \mathbb{E}_t \left[ (V_{\phi}(s_t) - \hat{V}_t)^2 \right],
\end{equation}
where \( \hat{V}_t \) represents the target value, and the total loss is:
\begin{equation}
L_{\text{total}} = L_{\text{actor}} + L_{\text{critic}} \quad \text{with} \quad \hat{V}_t = R_t + \gamma V_{\phi}(s_{t+1}).
\end{equation}

Double Deep Q-Network (DDQN) addresses the issue of overestimation in Q-learning by using two separate Q-networks to estimate the target Q-values: one for action selection and one for evaluation. The target value for updating the Q-network is:
\begin{equation}
y = r_t + \gamma Q_{\theta^-}(s_{t+1}, \arg\max_a Q_{\theta}(s_{t+1}, a)),
\end{equation}
where \( \theta^- \) refers to the target network parameters, and \( \arg\max_a Q_{\theta}(s_{t+1}, a) \) selects the action with the highest Q-value from the current Q-network \cite{van2016deep}. The loss function for DDQN is:
\begin{equation}
L_{\text{DDQN}} = \mathbb{E}_t \left[ \left( Q_{\theta}(s_t, a_t) - y \right)^2 \right],
\end{equation}
where \( y \) is the target value as defined above. The Q-values are updated iteratively with the following update rule:
\begin{equation}
Q_{\theta}(s_t, a_t) \leftarrow Q_{\theta}(s_t, a_t) + \alpha \left( y - Q_{\theta}(s_t, a_t) \right).
\end{equation}

Twin Delayed Deep Deterministic Policy Gradient (TD3) is an off-policy algorithm designed for continuous action spaces. TD3 improves the Deep Deterministic Policy Gradient (DDPG) by introducing three key modifications: target policy smoothing, delayed updates, and the use of two Q-functions for better value estimation \cite{fujimoto2018addressing}. The Q-function is updated using the Bellman equation with a target network:
\begin{align}
Q_1(s_t, a_t) &\leftarrow Q_1(s_t, a_t) + \alpha \big[ r_t + \gamma \min_{i=1,2} \nonumber\\
&\quad Q_{\theta^{'}}(s_{t+1}, a_{t+1}) - Q_1(s_t, a_t) \big],
\end{align}
where \( \theta^{'} \) refers to the parameters of the target networks, and \( \min_{i=1,2} \) refers to using the minimum value of the two Q-functions to mitigate overestimation bias. The policy is updated by maximizing the Q-function:
\begin{equation}
\theta_{\pi} \leftarrow \theta_{\pi} + \beta \nabla_{\theta_{\pi}} \mathbb{E}_t \left[ Q_1(s_t, \pi(s_t)) \right].
\end{equation}

The policy smoothing involves adding noise to the target policy during the Bellman backup:
\begin{equation}
a_{t+1} = \pi(s_{t+1}) + \mathcal{N}, \quad \text{where} \quad \mathcal{N} \sim \mathcal{N}(0, \sigma^2).
\end{equation}

\subsection{Curriculum Learning \& Quantum Curriculum Learning}

Curriculum Learning (CL) represents an intuitive approach to machine learning that emulates human cognitive development through progressive task complexity enhancement \cite{wang2021survey}. In the quantum machine learning domain, this paradigm evolves into Quantum Curriculum Learning (Q-CurL), introducing distinctive challenges and opportunities in quantum data and circuit optimization.

The theoretical framework is anchored in a task-based learning approach. Let \( \mathcal{H} = \{h : X \rightarrow Y\} \) represent the hypothesis set, where \( X \) denotes the input quantum state space and \( Y \) the output quantum state space. The primary objective is to find a hypothesis \( h \) that approximates the true underlying function \( f : X \rightarrow Y \) such that \( h(x) \approx f(x) \). The learning objective minimizes the expected risk defined over the joint probability distribution \( P(X,Y) \):
\begin{equation}
R(h) := \mathbb{E}_{(x,y) \sim P(X,Y)} [\ell(h(x), y)],
\end{equation}
where \( \ell : Y \times Y \rightarrow \mathbb{R} \) represents the loss function measuring the approximation error between the predicted output \( h(x) \) and the target output \( y \) \cite{tran2024quantum}.

In practice, the empirical risk is minimized using the observed dataset \( \mathcal{D} = \{(x_i, y_i)\}_{i=1}^{N} \):
\begin{equation}
\hat{R}(h) = \frac{1}{N} \sum_{i=1}^{N} \ell(h(x_i), y_i),
\end{equation}

The quantum curriculum comprises a sequence of auxiliary tasks \( \{T_1, \ldots, T_{M-1}\} \) strategically designed to enhance performance on a main task \( T_M \). Each task \( T_m \) is associated with a dataset \( \mathcal{D}_m = \{(x^{(m)}_i, y^{(m)}_i)\}_{i=1}^{N_m} \) drawn from a distribution \( P^{(m)}(X,Y) \) \cite{tran2024quantum}.

The curriculum weight \( c_{M,m} \), which quantifies the contribution of auxiliary task \( T_m \) to the main task \( T_M \). This weight is derived through a density ratio estimation:
\begin{equation}
c_{M,m} = \frac{1}{N_m} \sum_{i=1}^{N_m} \hat{r}_{\alpha}(x^{(m)}_i, y^{(m)}_i).
\end{equation}

The density ratio \( \hat{r}_{\alpha}(x,y) = \alpha^{\top} \phi(x,y) \) employs a linear model with a quantum kernel-based basis function:
\begin{equation}
\phi_l(x,y) = \text{Tr}[x x^{(M)}_l] \cdot \text{Tr}[y y^{(M)}_l].
\end{equation}

In quantum circuit learning, the objective is to tune the parameters (\( \theta \)) of a quantum circuit \( U(\theta) \) so that it closely approximates a target unitary operation \( V \), which is typically unknown. The goal is to adjust \( \theta \) such that the output of the quantum circuit matches the behavior of the target unitary as closely as possible. The optimization objective uses the Hilbert-Schmidt distance:
\begin{equation}
\mathcal{C}_{\text{HST}}(\theta) := 1 - \frac{1}{d^2} \left|\text{Tr}[V^{\dagger}U(\theta)]\right|^2,
\end{equation}
where \( d = 2^Q \) represents the Hilbert space dimension for a \( Q \)-qubit system.

For a training dataset \( \mathcal{D}_Q(N) = \{(|\psi\rangle_j, V|\psi\rangle_j)\}_{j=1}^{N} \), typically drawn from the Haar distribution, the empirical loss is defined as:
\begin{equation}
\mathcal{C}_{\mathcal{D}_Q(N)}(\theta) := 1 - \frac{1}{N} \sum_{j=1}^{N} |\langle\psi_j|V^{\dagger}U(\theta)|\psi_j\rangle|^2.
\end{equation}

The weights \( w_i \) are dynamically adjusted based on the curriculum weights \( c_{M,m} \), enabling a nuanced progression of task complexity.

\begin{figure*}
    \includegraphics[width=\linewidth]{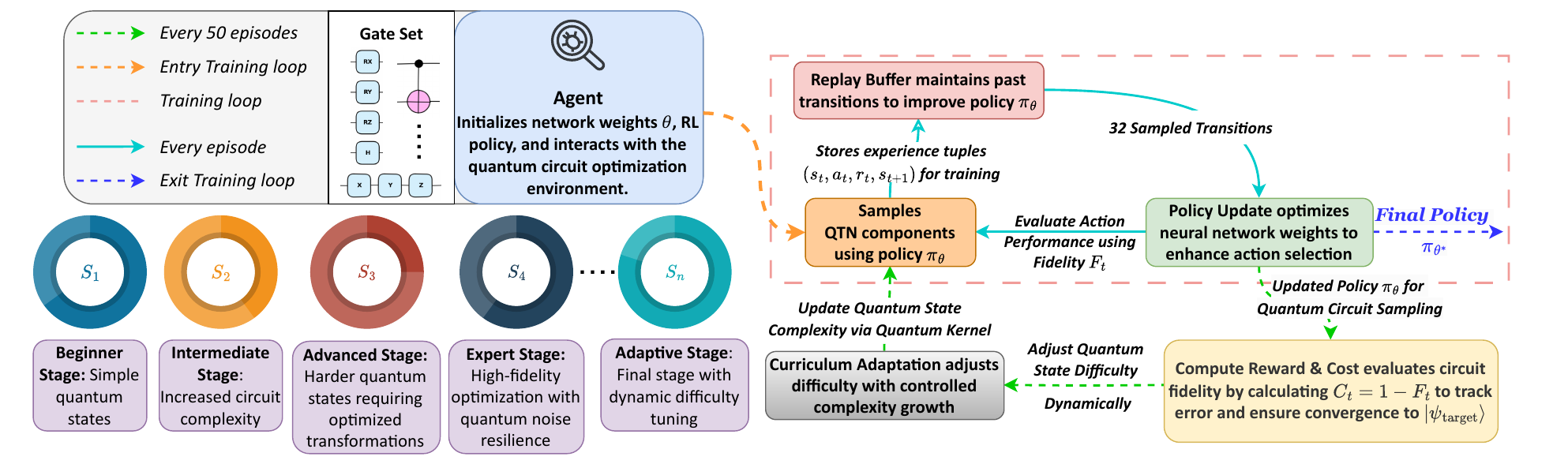}
    \label{fig:1}
    \vspace{-0.5cm}
    \caption{The proposed workflow illustrates the hybrid QRL framework for QAS, integrating curriculum learning to progressively optimize VQCs. The agent interacts with a quantum circuit optimization environment, sampling QTNs components while refining policies through RL. A replay buffer stores experience transitions to enhance learning, and circuit difficulty adapts dynamically based on fidelity evaluations, ensuring convergence to optimal quantum states.}
\end{figure*}

\subsection{Quantum Tensor Network}

QTNs are used to represent quantum states efficiently, providing a potential tool for QAS. A general quantum state expressed as a tensor network is:
\begin{equation}
|\psi\rangle = \sum_{i_1, i_2, \dots, i_k} T_{i_1, i_2, \dots, i_k} |i_1\rangle \otimes |i_2\rangle \otimes \dots \otimes |i_k\rangle,
\end{equation}
where \( T_{i_1, i_2, \dots, i_k} \) are tensor components, and \( |i_1\rangle, |i_2\rangle, \dots, |i_k\rangle \) are basis states. Tensor networks decompose a quantum state into smaller tensors, facilitating efficient manipulation and computation, especially for high-dimensional systems \cite{biamonte2019lectures}.

In the case of Variational Quantum Eigensolver (VQE), which uses quantum circuits for ground-state energy calculations, the loss function is often defined as:
\begin{equation}
\mathcal{L}_{\text{VQE}}(\theta) = \langle \psi(\theta) | \hat{H} | \psi(\theta) \rangle,
\end{equation}
where \( \hat{H} \) is the Hamiltonian of the system, and \( |\psi(\theta)\rangle \) is the quantum state generated by the parameterized quantum circuit. The QTNs model helps in minimizing this loss by efficiently representing quantum states and optimizing the variational parameters \cite{khan2023pre}.

\section{QAS-QTNs Framework\label{se3}}

\subsection{Problem Formulation: Quantum State Preparation and Targets for Simulations\label{se31}}

The Quantum Architecture Search with Quantum Tensor Networks (QAS-QTNs) framework provides an automated approach for designing efficient quantum circuits that prepare specific target quantum states. These quantum states serve as fundamental building blocks across various quantum computing applications, including quantum communication, error correction, and simulations. Our framework evaluation focuses on the preparation of maximally entangled quantum states, specifically Bell states and Greenberger–Horne–Zeilinger (GHZ) states \cite{nielsen2010quantum}.

The Bell state, which represents maximal two-qubit entanglement, is formally defined as:
\begin{equation}
|\Phi^+\rangle = \frac{1}{\sqrt{2}} \sum_{i=0}^1 |i\rangle|i\rangle.
\end{equation}

Using ZX-Calculus, we can represent this definition through a curved line (also known as a cup):
\begin{align}
\diagram{
    \begin{scope}[xscale=-1]
        \draw[ten] (-1/2,-2/2) -- (-1/2,2/2) -- (1/2,2/2) -- (1/2,-2/2) -- cycle;	
        \node {$\Omega$};
        \draw (0.5,0.75) -- (1,0.75);
        \draw (0.5,-0.75) -- (1,-0.75);
    \end{scope}	
}
&= \frac{1}{\sqrt{2}}\hspace{-0.125cm}
\diagram{
    \begin{scope}[xscale=-1]
        \draw (0.5,0.75) .. controls (-0.5,0.75) and (-0.5,-0.75) .. (0.5,-0.75);
        \draw (0.5,0.75) -- (1.25,0.75);
        \draw (0.5,-0.75) -- (1.25,-0.75);
    \end{scope}	
}
\end{align}

We extend this by defining \( \Omega(O) \) as the vectorisation of an operator \( O \), where \( \ket{\Omega(O)} = (O \otimes I) \ket{\Omega} \):
\begin{align}
\diagram{
    \begin{scope}[xscale=-1]
        \draw[ten] (-3/2,-2/2) -- (-3/2,2/2) -- (1/2,2/2) -- (1/2,-2/2) -- cycle;	
        \node at (-1/2,0) {$~\Omega(O)$};
        \draw (0.5,0.75) -- (1,0.75);
        \draw (0.5,-0.75) -- (1,-0.75);
    \end{scope}	
}
&= \frac{1}{\sqrt{2}}\hspace{-0.125cm}
\diagram{
    \begin{scope}[xscale=-1]
        \draw[white] (0,1.25) -- (0,0);
        \draw (0.5,0.75) .. controls (-0.5,0.75) and (-0.5,-0.75) .. (0.5,-0.75);
        \draw (0.5,0.75) -- (2,0.75);
        \draw (0.5,-0.75) -- (2,-0.75);	
        \draw[ten] (0.5,1.25) -- (0.5,0.25) -- (1.5,0.25) -- (1.5,1.25) -- cycle;
        \node at (1,0.75) {$O$};
    \end{scope}
}
\end{align}

This definition reveals that the Bell basis corresponds directly to the vectorisation of the Pauli operators:
\begin{equation}
\ket{\Phi^+}=\ket{\Omega(I)}, \quad \ket{\Phi^-}=\ket{\Omega(Z)},
\end{equation}
\begin{equation}
\ket{\Psi^+}=\ket{\Omega(X)}, \quad \ket{\Psi^-} \propto \ket{\Omega(Y)}.
\end{equation}

The Bell basis maintains an intrinsic connection to the Pauli operators, where the Euclidean inner product on Bell basis states corresponds to the Hilbert-Schmidt inner product on Paulis.\cite{bridgeman2017hand} Moving to three-qubit systems, the GHZ state represents maximal entanglement among three qubits:
\begin{equation}
|GHZ\rangle = \frac{1}{\sqrt{2}} \sum_{i=0}^1 |i\rangle|i\rangle|i\rangle.
\end{equation}

This can be elegantly represented using the following state diagram. Let:
\begin{align}
\begin{array}{c}
\begin{tikzpicture}[scale=.25]
    \draw (0,0)--(0,4.5) (0,-1) node {$j$} (0,5.2) node {$l$};
    \draw (-2,0)--(-2,2) (-2,-1) node {$i$};
    \draw (2,0)--(2,2) (2,-1) node {$k$};
    \filldraw[w] (-2,2)--(2,2)--(0,3.5)--(-2,2);
\end{tikzpicture}
\end{array}
&= \delta_{i,j,k,l},
\end{align}
where \( u = \mathbbm{1} \) and the top tensor \( T = \ket{+} \). This construction yields the state:
\begin{equation}    
\frac{\ket{0}^{\otimes N} + \ket{1}^{\otimes N}}{\sqrt{2}}.
\end{equation}

This quantum computational synthesis methodology's effectiveness is rigorously evaluated through two critical analytical indicators.

\subsubsection{State Generation Reliability}

This indicator captures the agent's ability to discover quantum circuit architectures capable of manifesting a predetermined quantum configuration. The metric is derived by calculating the ratio of successfully identified viable circuit designs against the total potential circuit configurations.

\begin{equation}
R_{\text{success}} = \frac{N_{\text{acceptable circuits}}}{N_{\text{total configurations}}}.
\end{equation}

A viable circuit is a quantum circuit design that creates a quantum state similar to the desired quantum representation.

\subsubsection{Optimization Potential Assessment}

This evaluation examines how well the methodology identifies functional circuit designs and builds configurations that achieve the highest computational efficiency. An optimized configuration reduces complexity by reducing the circuit depth and gate usage while accurately producing the desired quantum state.

\begin{equation}
R_{\text{optimal}} = \frac{N_{\text{optimal circuits}}}{N_{\text{total configurations}}}.
\end{equation}

This metric fundamentally assesses the ability of the computational synthesis approach to generate strategically engineered quantum circuit architectures.

\begin{figure}
\centering
\hspace{-20pt}
\begin{minipage}{\textwidth}
\Qcircuit @C=1em @R=.7em {
& \gate{H} & \ctrl{1} & \qw \\
& \qw & \targ & \qw
}
\end{minipage}
\hspace{20pt}
\begin{minipage}{\textwidth}
\Qcircuit @C=1em @R=.7em {
& \gate{H} & \ctrl{1} & \qw \\
& \qw & \targ & \ctrl{1} \\
& \qw & \qw & \targ
}
\end{minipage}
\caption{Quantum circuits for entanglement generation: (left) a Bell state (\(|\Phi^+\rangle\)) is created by applying a Hadamard gate to the first qubit, followed by a CNOT gate; (right) a GHZ state is generated by extending the Bell state with a second CNOT gate to entangle a third qubit.}
\vspace{-15pt}
\label{fig:2}
\end{figure}
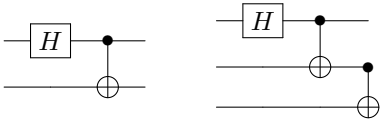

\subsection{Methodology \label{se32}}

\begin{algorithm}[t]
\caption{QAS-QTN Algorithm with Quantum Curriculum Learning}
\label{alg:qas_qtn}
\begin{algorithmic}[1]
\State \textbf{Input:} Target quantum state \( |\psi_{\text{target}}\rangle \), initial quantum state \( |\psi_{\text{in}}\rangle \), quantum circuit \( U(\vec{\theta}) \) with parameters \( \vec{\theta} \in \mathbb{R}^d \), action space \( \mathcal{A} \), number of episodes \( E \), curriculum stages \( \mathcal{S} = \{T_1, T_2, \dots, T_M\} \)
\State \textbf{Output:} Optimized quantum circuit \( U^*(\vec{\theta}) \)
\State Initialize RL agent \( \mathcal{A}_\text{agent} \) with empty quantum circuit model
\State Define reward function \( R_t \) and cost function \( C_t \) based on fidelity \( F_t \)
\State Initialize curriculum stages \( \mathcal{S} = \{T_1, T_2, \dots, T_M\} \) with associated datasets \( \mathcal{D}_m = \{(x^{(m)}_i, y^{(m)}_i)\}_{i=1}^{N_m} \) for each task \( T_m \)
\For {each curriculum stage \( s \in \mathcal{S} \)}
    \State Set current stage parameters: \( |\psi_{\text{target}}^{(s)}\rangle \), complexity level, and quantum state difficulty
    \For {each episode \( e = 1, 2, \dots, E \)}
        \State Set initial state \( |\psi_{\text{in}}\rangle \)
        \State Initialize time step \( t = 1 \)
        \While {not terminated (until \( t = T_e \))}
            \State Select action \( a_t \in \mathcal{A} \) according to policy \( \pi(a_t | s_t) \)
            \State Update quantum circuit: \( U(\vec{\theta}) \leftarrow U(\vec{\theta}) \cdot G(a_t) \)
            \State Calculate cost \( C_t = 1 - F_t = 1 - |\langle \psi_{\text{target}}^{(s)} | \psi_{\text{out}}(t) \rangle|^2 \)
            \State Compute reward \( R_t \) using cost function \( C_t \)
            \State Update policy \( \pi(a_t | s_t) \) using RL method (e.g., Q-learning, Policy Gradient)
            \State Update exploration rate \( \epsilon_t = \max(\epsilon_{\text{min}}, \epsilon_{\text{start}} \cdot \epsilon_{\text{decay}}^t) \)
            \State Store transition \( (s_t, a_t, r_t, s_{t+1}) \) in replay buffer \( \mathcal{D} \)
            \State \( t \leftarrow t + 1 \)
        \EndWhile
        \State \textbf{Curriculum Update:} Adjust quantum state difficulty or introduce noise in dataset \( \mathcal{D}_m \) based on curriculum weight \( c_{M,m} \)
        \State Dynamically adjust the curriculum weight: 
        \[
        c_{M,m} = \frac{1}{N_m} \sum_{i=1}^{N_m} \hat{r}_{\alpha}(x^{(m)}_i, y^{(m)}_i)
        \]
        where \( \hat{r}_{\alpha}(x,y) = \alpha^{\top} \phi(x,y) \) and the feature map \( \phi(x,y) \) uses a quantum kernel
    \EndFor
\EndFor
\State Return quantum circuit \( U^*(\vec{\theta}) \)
\end{algorithmic}
\end{algorithm}

The QAS-QTNs framework design and automation process involves the efficient representation of quantum states using QTNs, particularly matrix product states (MPS) and tree tensor networks (TTNs) \cite{rieser2023tensor}. The framework leverages RL to search for and optimize the quantum circuit, with each step mathematically framed to support the optimization of quantum state preparation, as illustrated in Fig. \ref{fig:1}. 

Let the target quantum state be represented by \( |\psi_{\text{target}}\rangle \), where we aim to prepare this state using a quantum circuit \( U(\vec{\theta}) \) parameterized by \( \vec{\theta} \in \mathbb{R}^d \) (with \( d \) being the number of parameters in the circuit). The quantum circuit's objective is to generate a quantum state \( |\psi_{\text{out}}\rangle = U(\vec{\theta}) |\psi_{\text{in}}\rangle \), where \( |\psi_{\text{in}}\rangle \) is the initial state of the quantum system (typically \( |0\rangle^{\otimes n} \), an \( n \)-qubit state). The preparation process is evaluated based on the overlap between the prepared state and the target state, quantified by the fidelity:
\begin{equation}
F = |\langle \psi_{\text{target}} | \psi_{\text{out}} \rangle|^2.
\end{equation}

The QTNs representation is employed to model the quantum states efficiently. Let the quantum state \( |\psi\rangle \) be represented as an MPS, which is a sequence of tensor operations on each qubit:
\begin{equation}
|\psi\rangle = \sum_{i_1,i_2,\dots,i_n} \text{Tr}\left( A^{[1] i_1} A^{[2] i_2} \dots A^{[n] i_n} \right) |i_1 i_2 \dots i_n\rangle,
\end{equation}
where \( A^{[k] i_k} \) are tensors associated with each qubit and \( i_k \) represents the possible states of the \( k \)-th qubit. For TTN, the representation involves a hierarchical tensor decomposition, reducing the complexity of describing entangled quantum states.

The RL agent builds quantum circuits iteratively, selecting gates to append based on the current quantum state. The agent operates in a discrete action space consisting of \( 3n + 2n^2 \) actions, where each action corresponds to either a CNOT gate (with control and target qubit positions) or a single-qubit rotation gate (with axis of rotation and qubit position).

Let the set of possible actions be \( \mathcal{A} = \{ a_1, a_2, \dots, a_{3n + 2n^2} \} \). The agent's objective is to find a sequence of actions \( \{ a_1, a_2, \dots, a_t \} \) that transforms the quantum state from \( |\psi_{\text{in}}\rangle \) to \( |\psi_{\text{out}}\rangle \), as represented by the corresponding quantum circuit \( U(\vec{\theta}) \).

To guide the agent's learning, a reward function \( R_t \) is defined at each time step \( t \) to evaluate the quality of the quantum state generated by the circuit:
\begin{equation}
R_t = \begin{cases}
5 & \text{if } C_t < \xi, \\
-5 & \text{if } t \geq T_e \text{ and } C_t \geq \xi, \\
\max \left( \frac{C_{t-1} - C_t}{C_{t-1} - C_{\text{min}}}, -1 \right) & \text{otherwise},
\end{cases}
\end{equation}
where \( C_t \) represents the cost function at time step \( t \), and \( C_{\text{min}} \) is the minimum possible cost for the target state. The threshold \( \xi \) is a user-defined precision level, and \( T_e \) is the maximum number of actions in an episode \cite{ostaszewski2021reinforcement}.
The cost function \( C_t \) quantifies the difference between the target quantum state \( |\psi_{\text{target}}\rangle \) and the current quantum state \( |\psi_{\text{out}}\rangle \), often defined as:
\begin{equation}
C_t = 1 - F_t = 1 - |\langle \psi_{\text{target}} | \psi_{\text{out}} \rangle|^2,
\end{equation}
where \( F_t \) is the fidelity at time step \( t \). This cost function reflects the accuracy of the quantum state preparation.

At each time step, the agent selects an action \( a_t \) from the action space \( \mathcal{A} \), which modifies the current quantum circuit by applying a gate \( G(a_t) \) to the quantum system. The agent's policy \( \pi(a_t | s_t) \) determines the probability of selecting action \( a_t \) given the current state \( s_t \), where the state is represented by the current quantum state encoded through tensor networks.

The policy is trained using any general RL, where the agent's objective is to maximize the cumulative reward over an episode. The reward function is used to update the agent's policy using temporal difference methods, such as Q-learning or Policy Gradient methods. Illegal actions are introduced to handle the combinatorial complexity of quantum circuit design and result in non-viable circuits, such as applying redundant gates that cancel out, thus making no progress in state preparation. These actions are pruned from the action space to reduce the search space and guide the agent toward more efficient solutions. The agent's action space is pruned based on the unitary properties of quantum gates. Redundant gates, such as consecutive CNOT gates acting on the same qubits, cancel each other out and, thus, are considered illegal.

To evaluate the performance of the quantum circuit designed by the QAS-QTN framework in a real-world application, we benchmark it against two other models. The first is a classical model, which serves as the baseline for comparison. The second is a quantum machine learning model, utilizing a random ansatz for the quantum portion while relying on classical optimization for the remaining components. The third model is the quantum circuit designed by QAS, which is optimized through the QAS-QTN framework using reinforcement learning (RL) to iteratively improve the quantum state preparation. Each of these models is evaluated for accuracy on the Iris dataset\cite{fisher1936use}, offering insights into the performance of the quantum model.

\section{Experiments\label{se4}}

\subsection{Experimental Setup \label{se41}}

The experimental framework is developed using PennyLane \cite{bergholm2018pennylane} and PyTorch \cite{paszke2019pytorch} to design and evaluate RL architectures for constructing Bell and GHZ quantum states. To ensure consistent performance monitoring, training episodes for the Bell state were divided into four distinct intervals. A similar segmentation approach is used for GHZ state training, enabling a standardized evaluation methodology across both quantum state preparations. Each experiment is conducted over 10,000 episodes, segmented as described.

\begin{table}[ht]
\caption{Hyperparameters Used in Training.}
\centering
\begin{tabular}{c|c}
\hline
\textbf{Hyperparameter} & \textbf{Value} \\ \hline
Batch size & 1,000 \\ 
Memory size & 15,000 \\ 
Neural network structure & 3 layers, 30 neurons per layer \\ 
Learning rate & 5e-4 with AdamW Optimizer \\ 
Target network update frequency & Every 50 episodes \\ 
Gamma & 5e-3 \\ \hline
\end{tabular}
\label{tab:hyperparameters}
\end{table}

\begin{table}[ht]
\centering
\caption{PER and Exploration Strategy Parameters.}
\begin{tabular}{c|c}
\hline
\textbf{Parameter} & \textbf{Value} \\ \hline
Epsilon decay rate & 0.99995 \\ 
Minimum epsilon & 5e-2 \\ 
Epsilon restart value & 1.0 \\ 
Alpha & 6e-1 \\ 
Initial beta & 4e-1 \\ 
Beta increment & 1e-3 per episode \\ \hline
\end{tabular}
\label{tab:exploration}
\end{table}

\begin{figure*}
    \centering
    \includegraphics[width=\linewidth]{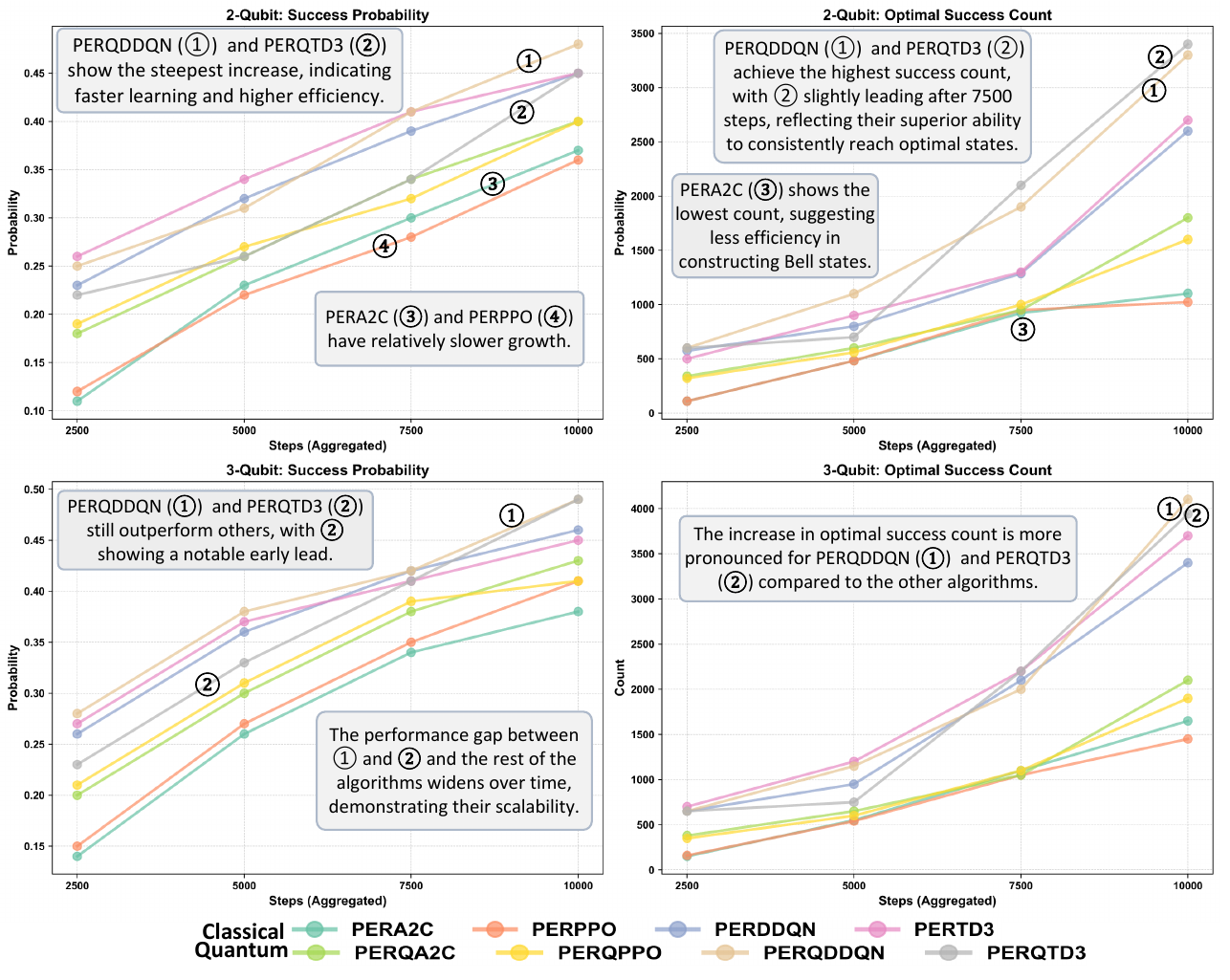}
    \vspace{-0.6cm}
    \caption{Comparison of RL algorithms in quantum environments, the plots illustrate the performance of multiple RL algorithms in constructing Bell and GHZ states across 2-qubit and 3-qubit quantum environments. Success Probability and Optimal Success Count metrics are tracked over 10,000 training steps. Hybrid quantum algorithms, such as PERQDDQN \& PERQTD3, exhibit superior learning efficiency and higher success rates in both environments, with the differences in performance becoming more pronounced in the more complex 3-qubit setting, highlighting the scalability and effectiveness of prioritization-based strategies in hybrid quantum RL tasks.}
    \label{fig:3}
\end{figure*}

To encourage exploration, the epsilon-greedy strategy is implemented, while Prioritized Experience Replay (PER) is incorporated into all RL models to improve learning efficiency. The specific parameters for both exploration and PER are outlined in Table \ref{tab:exploration}. At each time step \( t \), the action selection is determined as follows:
\begin{equation}
a_t = \begin{cases} 
\text{random action}, & \text{with probability } \epsilon_t \\
\arg\max_a Q(s_t, a), & \text{with probability } 1 - \epsilon_t
\end{cases}
\end{equation}
where \(a_t\) is the action taken at time step \(t\), \(s_t\) is the state at time step \(t\), \(Q(s_t, a)\) is the action-value function, and \(\epsilon_t\) is the exploration rate at time \(t\).

The exploration rate \(\epsilon_t\) decays over time according to the equation:
\begin{equation}
\epsilon_t = \max(\epsilon_{\text{min}}, \epsilon_{\text{start}} \cdot \epsilon_{\text{decay}}^t),
\end{equation}
where \(\epsilon_{\text{min}}\) is the minimum epsilon value, \(\epsilon_{\text{start}}\) is the initial epsilon value, \(\epsilon_{\text{decay}}\) is the decay rate, and \(t\) is the time step (or episode number). Quantum counterparts of the classical RL architectures, including DDQN, TD3, and PPO, are constructed with six logical qubits. 

\subsection{Results\label{se42}}
\subsubsection{Quantum Environment Results\label{se42.1}}

In the 2-qubit environment, as illustrated in Fig.~\ref{fig:3}, the algorithms show improvements in success probability and optimal success count as training progresses. PERDDQN and PERTD3 outperform the other algorithms, achieving final success probabilities of 0.42 and 0.41, respectively, at around 10,000 training steps. PERDDQN reaches about 2,400 optimal successes, followed closely by PERTD3 with approximately 2,300. PERA2C and PERPPO achieve success probabilities of 0.35 and 0.34, with lower optimal success counts of around 1,000 and 920, respectively. Notably, PERQDDQN exhibits a final success probability near 0.46, showing consistent growth throughout the training process and culminating in an optimal success count of approximately 3,000.

In the more complex 3-qubit environment, also shown in Fig.~\ref{fig:3}, the improvements across metrics are slower due to the increased complexity of the task. PERQDDQN and PERQTD3 demonstrate strong performance, achieving final success probabilities close to 0.47. These algorithms record the highest optimal success counts, with PERQDDQN reaching about 3,800 and PERQTD3 achieving approximately 3,600. PERDDQN and PERTD3 display success probabilities near 0.43 and 0.42, with optimal success counts around 3,100 and 3,200, respectively. PERA2C and PERPPO, with success probabilities of 0.36 and 0.39 and optimal success counts of approximately 1,500 and 1,300, struggle to scale effectively in this more complex environment.

The similar performance metrics of PERA2C and PERPPO, along with their quantum counterparts in both environments, likely arise because A2C is a special case of PPO with fewer optimizations. While PPO's clipping mechanism aims to improve policy stability, A2C's simpler implementation yields comparable results in environments with moderate complexity. In the 3-qubit case, PPO's adaptation mechanisms allow it to slightly outperform A2C. PERDDQN and PERTD3 exhibit comparable performance due to their shared underlying structure based on the Deep Deterministic Policy Gradient (DDPG) algorithm. Both approaches utilize a combination of deterministic policy gradients and actor-critic methods to optimize their decision-making processes. The key similarity lies in how they leverage this framework for continuous control tasks, resulting in similar outcomes in terms of success probability and optimal success count.

\subsubsection{Iris Dataset Classification Results\label{se42.2}}
In the experimental quantum machine learning environment for real-world applications, the classical model, which incorporates a single simple classical layer, achieves an accuracy of 72.11\% on the Iris dataset. When a single quantum layer with a random ansatz is introduced, consisting of 4 qubits and a depth of 3, the accuracy improves to 78.82\%. In contrast, the quantum model designed by QAS, which incorporates a fully optimized quantum circuit through the QAS-QTN framework, achieves an improvement, reaching an accuracy of 90.33\% with qubit system size of 4 and a circuit depth size of 3 emphasizing the performance gain that can be realized by utilizing a QAS-QTN optimized ansatz within the quantum layer, demonstrating its utility within the real-world applications in quantum machine learning tasks.

\section{Discussion \& Conclusion\label{se5}}
The proposed QAS-QTNs framework demonstrates the significant potential in automating quantum circuit design by integrating hybrid QRL with quantum curriculum learning strategies, leading to improved convergence rates and higher-quality quantum circuit designs compared to classical approaches. The experimental results, particularly in preparing Bell and GHZ states, show that quantum-enhanced algorithms like PERQDDQN and PERQTD3 outperform classical RL algorithms in both success probability and optimal success count, especially as the complexity of the task increases. This suggests that our framework effectively scales with the number of qubits, maintaining high performance even in the more complex 3-qubit GHZ state preparation. The use of quantum curriculum learning played a crucial role in managing task complexity, allowing the agent to progressively tackle more challenging quantum circuit optimization tasks. By systematically exploring the design space and pruning illegal or redundant actions, the agent efficiently discovered optimized circuits with reduced depth and gate counts, which is essential for practical implementations on NISQ devices with limited coherence times and gate fidelities. While the results are promising, future work should address challenges such as testing on actual quantum hardware to account for noise and decoherence, scaling to larger quantum systems, and further reducing the action space to enhance learning efficiency. Nonetheless, our work bridges the gap between QRL and quantum circuit optimization. Also, it sets the stage for potential advances in designing efficient circuits for quantum computing applications, moving us closer to realizing practical quantum algorithms capable of solving problems beyond the reach of classical computers.

\section*{Acknowledgment}
This work was supported in part by the NYUAD Center for Quantum and Topological Systems (CQTS), funded by Tamkeen under the NYUAD Research Institute grant CG008.
\bibliographystyle{IEEEtran}

\bibliography{refs}
\end{document}